\documentclass{aa}
\usepackage{graphicx}
\usepackage{txfonts}

\begin{document}
   
   \title{Stochastic particle acceleration and synchrotron self--Compton radiation in
          TeV blazars.
         }
      
   \author{Katarzy\'nski K. \inst{1,2}, 
           Ghisellini G.    \inst{1},
           Mastichiadis A.  \inst{3},
	   Tavecchio F.     \inst{1},
           Maraschi L.      \inst{1}
	  }

   \offprints{Krzysztof Katarzy\'nski \\kat@astro.uni.torun.pl}

   \institute{Osservatorio Astronomico di Brera, via Bianchi 46, Merate and via Brera 28, Milano, Italy  \and 
              Toru\'n Centre for Astronomy, Nicolaus Copernicus University, ul. Gagarina 11, PL-87100 Toru\'n, Poland \and
              Dept. of Physics, Univ. of Athens, Panepistimiopolis, 15784 Zografos, Athens 
             }

   \date{Received 8 September 2005 / Accepted 22 February 2006}

           \abstract{}
            {We analyse the influence of the stochastic particle acceleration for the evolution of the
             electron spectrum. We assume that all investigated spectra are generated inside a spherical,
             homogeneous source and also analyse the synchrotron and inverse Compton emission generated 
             by such an object. 
            }
            {The stochastic acceleration is treated as the diffusion of the particle 
             momentum and is described by the momentum--diffusion equation. We investigate the stationary
             and time dependent solutions of the equation for several different evolutionary scenarios.
             The scenarios are divided into two general classes. First, we analyse a few cases 
             without injection or escape of the particles during the evolution. Then we investigate
             the scenarios where we assume continuous injection and simultaneous escape of the 
             particles.
            }
            {In the case of no injection and escape the acceleration process, competing with the 
             radiative cooling, only modifies the initial particle spectrum. The competition leads 
             to a thermal or quasi--thermal distribution of the particle energy. In the case of
             the injection and simultaneous escape the resulting spectra depend mostly on the energy 
             distribution of the injected particles. In the simplest case, where the particles are 
             injected at the lowest possible energies, the competition between the acceleration and the
             escape forms a power--law energy distribution. We apply our modeling to the high energy 
             activity of the blazar Mrk~501 observed in April 1997. Calculating the evolution of 
             the electron spectrum self--consistently we can reproduce the observed spectra well 
             with a number of free parameters that is comparable to or less than in the ``classic 
             stationary" one--zone synchrotron self--Compton scenario.
            \keywords{Radiation mechanisms: non--thermal -- Galaxies:
             active -- BL Lacertae objects: individual: Mrk~501}
            }
            {}
             
\titlerunning{Stochastic particle acceleration in TeV blazars.}
\authorrunning{Katarzy\'nski et al.}   
\maketitle

\section{Introduction}

The overall emission from blazars (e.g. Urry \& Padovani \cite{Urry}) is
successfully explained as the result of synchrotron and
inverse--Compton emission by relativistic electrons within a
relativistic jet (e.g. Ghisellini et al. \cite{Ghisellini98}; for a different view
see e.g. Mucke \cite{Mucke}). Most of the models usually applied to
describe the emission adopt a phenomenological view, assuming that
some (usually unspecified) mechanism is able to produce the electron
distribution that is subsequently injected into the emission region (e.g.
Mastichiadis \& Kirk \cite{Mastichiadis97},
Chiaberge \& Ghisellini \cite{Chiaberge99}, 
Kataoka et al. \cite{Kataoka00},
Moderski et al. \cite{Moderski03}).
Two processes are usually considered as a possible source of the
relativistic particles: the first--order Fermi acceleration at a shock
front and the second--order Fermi acceleration by a plasma turbulence
in e.g. a downstream region of the shock.

In the first--order process, the particle crossing the shock discontinuity 
gains an amount of energy that is proportional to the shock velocity. 
Behind the shock the 
particle may be scattered back by magnetic inhomogeneities, and as a result 
it crosses the shock many times, each time gaining energy. This scenario is
well known as a relatively efficient acceleration process that provides a 
power--law energy spectrum (e.g. Bell \cite{Bell78}, Blandford  \& Ostriker 
\cite{Blandford78}, Drury \cite{Drury83}, Blandford \& Eichler \cite{Blandford87},
Jones \& Ellison \cite{Jones91}). 
This power--law spectrum is generated by a kind of competition between the 
acceleration and the escape of the particles from the shock region. 
In a simple approach, the shock acceleration may be described in
a deterministic way, where the efficiency of the process is described by
the characteristic acceleration time ($t_{\rm acc}$, e.g. Kirk et al. 
\cite{Kirk98}). 
With this approach the particle energy gain per time unit is precisely defined. 

The second--order Fermi acceleration assumes reflection of the charged particles 
by a magnetized cloud or, in more sophisticated cases, by the magnetic inhomogeneities 
or plasma waves. The particle may gain or lose energy depending on whether the
``mirror'' is approaching or receding. However, the probability of the head--on
collisions is higher than for the rear--on reflections. Therefore, on average,
the particle can gain energy. The main energy gain per bounce is proportional to the 
square of the mirror velocity. However, the net energy gain depends not only on 
the mirror velocity but also on the scattering rate. 
Note that the net energy gain for 
the rear--on collisions is larger than the energy losses (e.g. Longair 
\cite{Longair92}). The second--order Fermi acceleration is usually treated as 
a stochastic process and described as the diffusion of the particle energy.
This process has been successfully applied in many different astrophysical sources
(e.g. Eilek \& Henriksen \cite{Eilek84}, Schlickeiser \cite{Schlickeiser84} \& 
\cite{Schlickeiser89}, Dung \& Petrosian \cite{Dung94}, Miller \& Roberts 
\cite{Miller95}, Dermer et al. \cite{Dermer96}, Petrosian \& Liu 
\cite{Petrosian04}). 
However, the efficiency of the second--order process was quite frequently 
questioned in comparison with the first--order process. Therefore the turbulent 
acceleration was frequently neglected when investigating the electron spectrum 
evolution in extragalactic jets.

The numerical simulations performed recently by Virtanen \& Vaino (\cite{Virtanen05}) 
show that the efficiency of the second--order Fermi acceleration may be comparable to
the efficiency of the shock acceleration. Therefore, the turbulent acceleration may
significantly affect the evolution of the particle spectrum. 
Moreover, the acceleration process may be quite complex. 
For example, in the first step, the particle may be efficiently 
accelerated at the shock front. 
After the escape into the downstream region of the shock, the particle is still 
accelerated by turbulent plasma waves. 
Therefore, the particle energy may increase enough to let the particle re--enter 
the shock acceleration region. 
In such a case the energy spectrum formed by the stochastic process may be 
re--accelerated by the shock.

For the sake of simplicity, we do not describe the acceleration
scenario in details. 
We assume that the acceleration process has a stochastic nature and
describe it as the diffusion of the particle momentum. 
This assumption does not mean that we consider only the second--order Fermi acceleration. 
We assume that the acceleration scenario is quite complex as in the example described above. 
Therefore the particle energy gain should be treated as a stochastic process and 
described by the diffusion. 
The efficiency of the diffusion is a free parameter in our model. 
We analyse the influence of this process in detail for the evolution of the particle spectrum. 
Moreover, we also analyse the evolution of the emission generated by the accelerated 
electrons inside a homogeneous spherical source. 
The electrons are producing the synchrotron and inverse Compton (hereafter IC) radiation. 
The synchrotron radiation field inside the source is scattered by the same population of  
electrons that produces this radiation. 
This is the synchrotron self--Compton process (hereafter SSC). 
This simple model may have direct application to many astrophysical sources and in particular 
to those BL Lac objects, that produce the very high energetic gamma rays observed 
up to the TeV band (e.g., Catanese et al. \cite{Catanese97}, Maraschi et al. 
\cite{Maraschi99}, Krawczynski et al. \cite{Krawczynski01}).

In all the simulations performed in this work, the acceleration process
is fully compensated for at some energy by the radiative cooling.  This
compensation leads to the stationary distribution of the particle
energy.  In a first approach we analyse a few evolutionary scenarios,
where we neglect the possible injection or escape of the particles
during the evolution of the source.  Then we investigate a few cases
where we introduce the continuous injection and escape of the
particles from the accelerating region.  Our results can be used to
investigate the origin of the variability of Mrk~501 observed in April
1997 (Pian et al. \cite{Pian98}, Djannati--Atai et
al. \cite{Djannati99}).  We will show that our model successfully
explains the different high states observed in this source, without
having to use any more free parameters than the most commonly used
one--zone SSC model.

\section{The momentum--diffusion equation}

The main difference between this paper and the previous works about modeling 
the TeV blazar emission is in the description of the particle 
acceleration process. We describe it as diffusion in the particle momentum 
space, where the evolution of the isotropic, homogeneous phase--space density 
($f$) is described by the momentum--diffusion equation:
\begin{equation}
\frac{\partial f(p,t)} {\partial t} = \frac{1}{p^2} \frac{\partial}{\partial p}
\left[ p^2 D(p,t) \frac{\partial f(p,t)}{\partial p} \right]
\end{equation}
where $p=\beta\gamma$ is the dimensionless particle momentum, $D(p,t)$ the 
momentum--diffusion coefficient, $\gamma$ the particle Lorentz factor 
($p=\sqrt{\gamma^2-1)}$), and $\beta$ the particle velocity in units of $c$
(e.g. Borovsky \& Eilek \cite{Borovsky86}). The particle number density ($N$) 
is directly related to the phase-space density:
\begin{equation}
N(p,t) = 4 \pi p^2 f(p,t).
\end{equation}
Therefore, we can rewrite this equation:
\begin{equation}
\frac{\partial N(p,t)} {\partial t} = \frac{\partial}{\partial p}
\left[-A(p,t)~N(p,t) + D(p,t) \frac{\partial N(p,t)}{\partial p} \right],
\end{equation}
where
\begin{equation}
A(p,t) = \frac{2}{p} D(p,t)
\end{equation}
describes the average efficiency of the acceleration process. In the case of 
ultra--relativistic particles ($\beta \simeq 1$), the particle momentum becomes 
equivalent to the particle Lorentz factor ($p \equiv \gamma$). 
Therefore we can rewrite the equation in the form:
\begin{equation}
\frac{\partial N(\gamma,t)} {\partial t} = \frac{\partial}{\partial \gamma}
\left[-A(\gamma,t)~N(\gamma,t) + D(\gamma,t) \frac{\partial N(\gamma,t)}{\partial 
\gamma} \right],
\end{equation}
(e.g. Brunetti \cite{Brunetti04}). In order to describe the radiative cooling 
of the particles and their possible escape or injection, we have to introduce 
three more terms
\begin{eqnarray}
\frac{\partial N(\gamma,t)} {\partial t} & = & \frac{\partial}{\partial \gamma}
\left[\left\{C(\gamma,t)-A(\gamma,t)\right\}N(\gamma,t) + D(\gamma,t) \frac{\partial 
N(\gamma,t)}{\partial \gamma} \right]\nonumber \\
&-& E(\gamma,t) + Q(\gamma,t).
\label{equ_kinetic}
\end{eqnarray}
The $C(\gamma, t)$ parameter describes the synchrotron and IC cooling of the particles
\begin{equation}
C(\gamma, t) = \frac{4}{3} \frac{\sigma_T c}{m_e c^2} \gamma^2 \left[ U_B(t)+U_{\rm rad}(\gamma, t) \right],
\end{equation}
where $U_B$ and $U_{\rm rad}$ are magnetic and radiation field energy densities,
respectively, $m_e$ is the electron rest mass, and the $\sigma_T$ is the Thomson
cross section. The escape of the particles is described by the $E(\gamma,t)$ term. 
For the sake of simplicity, in all calculations presented in this work we assume that
the escape term is independent of energy:
\begin{equation}
E(\gamma,t) = \frac{N(\gamma,t)}{t_{\rm esc}},
\end{equation}
characterized by a constant escape time ($t_{\rm esc}$) related to the source size. 
The injection of the particles is described by $Q(\gamma, t)$ and we define
this function separately for each test. 

Finally, to complete the description of the equation, we have to specify the diffusion 
coefficient. In all our tests we assume a Fermi--like acceleration process and define the 
coefficient as a time--independent power--law function of the particle energy
\begin{equation}
D(\gamma) = \frac{\chi}{2} \gamma^2,
\end{equation}
where $\chi = 1/t_{\rm acc}$ represents the efficiency of the acceleration process 
described by a characteristic acceleration time ($t_{\rm acc}$). The acceleration
time is a free parameter in our model. The above definition of the diffusion process 
gives the following relation for the acceleration term
\begin{equation}
A(\gamma) = \frac{\gamma}{t_{\rm acc}}.
\end{equation}
Note that this acceleration term describes the particle energy gain per time unit 
precisely and is identical to the terms used by other authors (e.g. Kirk et al. 
\cite{Kirk98}) in those simulations where the acceleration process is described in
a deterministic way. 
This helps to directly compare our test with the other simulations.

\section{A quasi--thermal stationary solution}
\label{sec_stationary}

In the first approach we analyse a stationary solution ($\dot{N}=0$) of Eq. 
\ref{equ_kinetic} in the case of the simplified cooling ($C=C_0\gamma^2$, i.e.
no synchrotron self--absorption is considered, and the 
IC cooling is assumed to occur entirely in the Thomson regime)
and without injection and escape of the particles during the evolution of the system 
\begin{equation}
\frac{\partial}{\partial \gamma}\left[\left\{C(\gamma)-A(\gamma)\right\}N(\gamma) + 
D(\gamma) \frac{\partial N(\gamma)}{\partial \gamma} \right] = 0.
\end{equation}
Note that the above assumption requires that the initial energy distribution 
$N(\gamma, t=0) = N_{\rm ini} (\gamma) \neq 0$. 
According to Chang \& Cooper 
(\cite{Chang70}), the general solution of this equation has the form
\begin{equation}
N(\gamma) = x \exp \left[ -\int^{\gamma} \frac{C(\gamma')-A(\gamma')}{D(\gamma')} d \gamma'\right],
\end{equation}
where $x$ is an integration constant. 
For our particular assumption about the 
diffusion process, the stationary solution is an ultrarelativistic Maxwellian
\begin{eqnarray}
N(\gamma) & = & x~\exp\left[ -\int^{\gamma}_1 
                  \frac{2 t_{\rm acc} \left(C_0\gamma'^2 - \frac{\gamma'}{t_{\rm acc}}\right)}{\gamma'^2} d \gamma' \right] \nonumber \\
          & = & x~\gamma^2 \exp[-2 C_0 t_{\rm a   cc} (\gamma-1)].
\end{eqnarray}
The maximum of this function appears at the equilibrium energy ($\gamma_{\rm e}$), where 
the acceleration process is fully compensated for by the cooling. If we describe the cooling 
using the characteristic cooling time
\begin{equation}
t_{\rm cool}(\gamma) = \frac{1}{C_0 \gamma},
\end{equation}
then at the equilibrium we have $t_{\rm cool}(\gamma_{\rm e}) = t_{\rm acc}$ that gives
\begin{equation}
\gamma_{\rm e} = \frac{1}{t_{\rm acc} C_0}.
\end{equation}

This simple solution shows a fundamental difference between the evolution where the 
acceleration is treated as a diffusion process and the acceleration described only 
by the deterministic acceleration term. 
In the case of the deterministic acceleration,
the system will try to reach the stationary state accelerating all the particles to 
the equilibrium energy and to build a monoenergetic population of the particles at 
this energy (``pile-up'').

Let us analyse now a more general case where the diffusion and cooling coefficients 
are defined with power law functions ($D=D_0 \gamma^{a}$ and $C=C_0 \gamma^{b}$). 
The stationary solution for this case is given by
\begin{equation}
N(\gamma) = x~\gamma^2 \exp \left[ \frac{C_0 \left(\gamma^{b+1-a}-1\right)}{D_0(a-1-b)} \right],
\end{equation}
where the power law part of the function always has the same slope ($\gamma^2$), but the
shape of the exponential part depends on the relation between the particular properties 
of the diffusion and cooling processes. 
Therefore in some physical scenarios, we expect the formation of a 
quasi--Maxwellian distribution, not a perfect Maxwellian. 
It should also be mentioned here that in the stationary spectrum
only the value of the constant $x$ depends on the initial energy distribution.  
If we know the total number of the particles in the system
\begin{equation}
N_{\rm tot} = \int^{\infty}_{1} N_{\rm ini}(\gamma) d \gamma,
\end{equation}
then we can easily find the value of this constant
\begin{equation}
x = \frac{N_{\rm tot}}{\int^{\infty}_1 \gamma^2 \exp 
    \left[ \frac{C_0 \left(\gamma^{b+1-a}-1\right)}{D_0(a-1-b)} \right] d \gamma}.
\end{equation}

The most important conclusion that appears from the analysis of the stationary 
solution is that the acceleration of the particles described as the diffusion process 
when competing with the radiative cooling may lead to a thermal or quasi--thermal 
particle distribution. 
This effect was showed for the first time by Schlickeiser (\cite{Schlickeiser84}) 
in a somewhat more complex way. 
The temperature of this thermal or quasi--thermal distribution is associated to 
the value of the equilibrium energy. 
The shape of the stationary spectrum is independent of the initial distribution,
which only determines the absolute level of the stationary solution.

Recently Saug\'e \& Henri (\cite{Sauge04}) proposed a scenario where the quasi--Maxwellian 
distribution of the electrons obtained from the acceleration, described as a diffusion 
process, is injected into a spherical region where the particles are producing the SSC
emission. 
They successfully apply this model to the activity of Mrk~501 observed
in April 1997 (Pian et al. \cite{Pian98}, Djannati--Atai et al. \cite{Djannati99}).
However, they did not study the acceleration process by assuming an arbitrary spectrum
of the injected particles.

A log--parabolic energy distribution, which is similar to a 
thermal distribution, has been proposed by Massaro et al. (\cite{Massaro04}) in order to 
explain the X--ray and TeV spectra of Mrk 421. 
However, this log--parabolic distribution
was obtained by assuming some specific nature of the acceleration process.

\section{Numerical approach} 

In several simple cases, it is possible to find an analytic, time--dependent
solution to Eq. \ref{equ_kinetic}. 
However, in the case of the SSC emission, where the radiation field is interacting 
with the particles that have generated this radiation, this partial differential 
equation becomes a partial differential integral equation. 
This poses a large problem for a attempt to analytically solve this equation. 
Therefore, in our simulations we use
the numerical method to find the time--dependent solutions.

The partial differential equation that describes the evolution of the particle
energy is a Fokker--Planck type equation. 
We use a very useful numerical differentiation scheme in our computations proposed 
for such type of equations by Chang \& Cooper (\cite{Chang70}). 
This method assumes the forward differentiation 
in time and the centred differentiation in the energy that is continuously 
transformed into the forward differentiation (see e.g. Press et al. \cite{Press89} 
for more information about the numerical differentiation). 
The transformation is correlated with the current position of the equilibrium 
between the acceleration and cooling. 
This is the big advantage of this method that provides stability for the numerical 
solution in the presence of equilibrium. 
Note that a constant differentiation method
in the energy range (e.g. centred) can be used effectively only in the case of a 
monotonic evolution of the spectrum towards lower ($t_{\rm acc} \gg t_{\rm cool}$) 
or higher ($t_{\rm acc} \ll t_{\rm cool}$) energies during the evolution of the 
system (e.g. Chiaberge \& Ghisellini \cite{Chiaberge99}, Moderski et al. 
\cite{Moderski03}). 
Moreover, the constant differentiation method also limits the shape of the 
initial spectrum or the injection function. 
The relatively ``sharp'' initial/injected function (e.g. Dirac delta function) may 
cause a large numerical diffusion, which can be reduced if we increase the 
number of the mesh points at the cost of calculation efficiency. 
However, this
increase may quickly lead to large numerical instabilities (see e.g. Press 
et al. \cite{Press89}). The method that we use is free from these problems. 
It is an implicit method, which means that it is necessary to solve 
a system of linear equations in order to obtain the solution. 
This provides additional stability for the numerical computations. 
The system of equations forms a tridiagonal matrix that can be easily 
solved numerically (e.g. Press et al. \cite{Press89}). 
In our computations we have slightly modified the 
originally proposed method by using logarithmic steps in the energy range. 
This helps to greatly reduce the number of mesh points necessary to obtain 
a precise result.

The additional advantage of this differentiation scheme is that the method 
provides conservation of the total particle number in the absence of external 
sources or sinks. 
Therefore, the result of numerical computation can be compared, for example, 
with the analytic stationary solutions discussed in the previous section. 
For the typical number of the time and energy mesh points used in 
our computations, we obtained conservation of the total particle number 
with precision better than 0.01$\%$. 
Moreover, we tested our numerical code with some basic time--dependent 
analytical solutions and found very good agreement.

\section{A simple evolution of the energy spectrum}
\label{sec_elecspec}

In this section we present a few simple cases the energy spectrum that
evolves in the case of the stochastic acceleration. 
In all tests we assumed some initial energy distribution and chose a constant 
efficiency for the acceleration process ($A = \gamma / t_{\rm acc}$) that in 
the competition with a constant cooling ($C=C_0 \gamma^2$) 
provides equilibrium at an energy equivalent to $\gamma_{\rm e} = 10^4$. 

\begin{figure}[p]
\resizebox{\hsize}{!}{\includegraphics{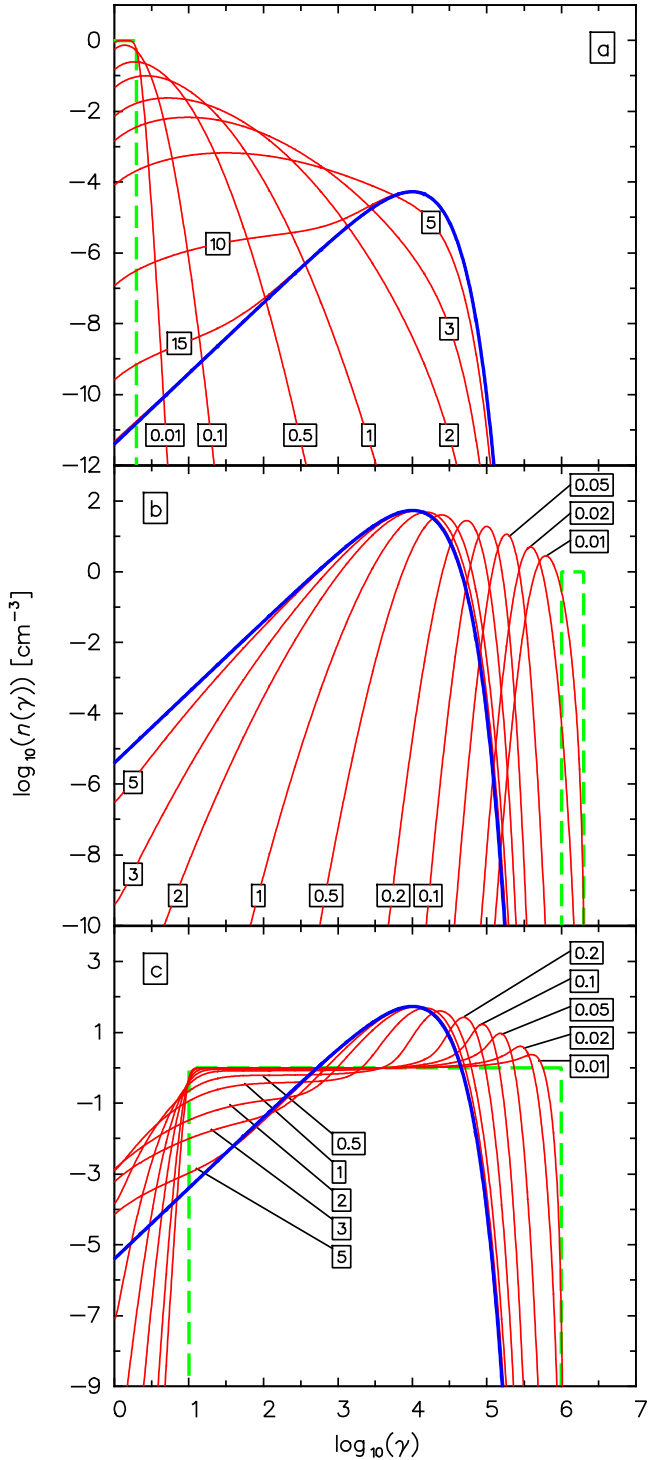}}
\caption{An example of the evolution of the electron spectrum in the case of no injection and
         escape of the particles. The upper panel (a) shows the evolution of the 
         initial spectrum (bold dashed line) where the particle energy ($1\le \gamma 
         \le 2$) is significantly lower than the equilibrium energy ($\gamma_{\rm e} = 
         10^4$). The opposite case, where the initial energy ($10^6 \le \gamma \le 2 
         \times 10^6$) is higher than the equilibrium energy, is presented in the middle 
         panel (b). The lower panel (c) shows a case where the lower energy in the 
         initial spectrum is lower than $\gamma_{\rm e}$, but the upper initial energy
         is higher than the equilibrium energy ($10 \le \gamma \le 10^6$). The 
         stationary Maxwellian spectrum on all figures is indicated by a bold solid 
         line. The numbers indicate the evolution time in units of $t_{\rm acc}$.
        }
\label{fig_max}        
\end{figure}

\subsection{No injection and escape}
\label{sec_elecspec_noie}

In the first three tests we neglected injection and escape of the particles during the
evolution of the system, and we analysed only the evolution of the initial spectrum.
\begin{itemize}
\item[$\bullet$]
     {First, we assume a very narrow initial distribution ($1\le \gamma \le 2$)
      with an average energy of the particles significantly less than the 
      equilibrium energy. The density of the particles is constant over the 
      whole energy range of the initial spectrum. In this particular
      case $t_{\rm acc} \ll t_{\rm cool}$ for the initial energy of the 
      particles. Therefore, the acceleration process tries to increase the 
      energy of all the particles. In the case of the stochastic acceleration,
      some of the particles may reach the equilibrium energy or even a higher 
      energy after an infinitesimally short time of the system evolution 
      ($t_0+\Delta t$). However, the number of the particles that can gain this 
      energy is negligibly small in comparison to the initial number of particles. 
      The system needs more time to accumulate a significant number of the 
      particles around the equilibrium energy. The simulation presented 
      in Fig.~\ref{fig_max}a shows a systematic decrease in the particle
      density for the initial energy range and a simultaneous increase in 
      the density around the equilibrium energy. In this particular case
      the densities in these two energy ranges become comparable after
      the evolution time of $t_{\rm evo} \gtrsim 5 t_{\rm acc}$. After a longer
      time ($t_{\rm evo} \gtrsim 20~t_{\rm acc}$), the acceleration will 
      significantly decrease the density of the particles with the 
      initial energy, accumulating a dominant part of the particles around 
      the equilibrium energy and reaching the stationary Maxwellian distribution.
      In the stationary state all of the energy acquired by the particles in
      the acceleration process is radiated away. 
      Note that the deterministic acceleration in this particular
      test could shift the initial spectrum towards the equilibrium
      energy without significantly modifying the shape as long as 
      the average energy of the particles in the system is significantly 
      less than $\gamma_{\rm e}$. Close to the equilibrium any initial 
      spectrum should be transformed by the deterministic acceleration
      into a quasi--monoenergetic distribution. Finally, the deterministic 
      acceleration appears less efficient than the stochastic acceleration.
      For the same value of the acceleration time, this process requires
      at least twice as much time to build a particle density that is comparable
      to the density provided by the stochastic acceleration around the
      equilibrium energy.
     }
\item[$\bullet$]
     {In the second test (Fig.~\ref{fig_max}b) we again assume a very narrow 
      initial distribution, but now the average energy of the particles 
      is significantly higher ($10^6 \le \gamma \le 2 \times 10^6$) than the 
      equilibrium energy. At the beginning the cooling process dominates the 
      evolution of the particle spectrum ($t_{\rm cool} \ll t_{\rm acc}$). 
      Therefore the initial distribution moves towards the equilibrium energy. 
      Note that in the case of a deterministic acceleration this particular 
      initial spectrum could be transformed into a quasi--monoenergetic distribution. 
      This is related to the fact that the cooling time is shorter for the 
      particles with the highest energy. In our test, instead of the monoenergetic 
      distribution, we obtain a kind of peaked spectrum that is more extended 
      in the energy range than the initial spectrum. However, the peak energy of 
      this distribution evolves in the same way as the energy of the 
      monoenergetic distribution in the case of no diffusion. When the
      peak energy reaches $\gamma_{\rm e}$, the shift in the peak ceases
      because the cooling process becomes fully compensated for by the acceleration.
      On the other hand, after this moment in the evolution, the left part of the 
      peaked distribution starts to change slope from an exponential 
      to a power law. When the index of this power law becomes equal to 2, this
      evolution stops and the formation of the stationary Maxwellian
      distribution is completed.
     }     
\item[$\bullet$]
     {In the next test (Fig.~\ref{fig_max}c), the assumed initial spectrum is relatively broad 
     ($10 \le \gamma \le 10^6$). The initial minimal energy of the particles is significantly 
      less than the equilibrium energy, but the maximal energy is significantly higher 
      than $\gamma_{\rm e}$. For the sake of simplicity we also assume a constant 
      particle density in this energy range. The specific assumption about the
      initial spectrum makes this case more complex than the previous ones.
      However, the evolution in this case in some specific energy ranges is very
      similar to the cases just discussed above. The change in the distribution 
      before the minimal initial energy is similar to the evolution presented in our 
      second test below the peak energy. Above the minimal energy, but below the 
      equilibrium, the changes in the spectrum are similar to the evolution presented 
      in our first test. Finally, above $\gamma_{\rm e}$ the evolution is analogous to 
      the variations presented in our second test, above the equilibrium.
     }          
\end{itemize}

The tests described above show that indeed the stationary state is almost independent 
of the initial distribution in the absence of the injection and escape of the 
particles. Note that a constant escape of the particles ($E(\gamma, t) = N(\gamma,t)) / 
t_{\rm esc}$) could produce a systematic decrease in the particle density proportional 
to a factor $\exp(-t/t_{\rm esc})$ without modifying the spectral shape (e.g. 
Schlickeiser (\cite{Schlickeiser84}). However, the escape and a simultaneous
continuous injection are able to completely change the evolution and the resulting 
stationary spectrum. We discuss such scenarios in the next subsection.

\subsection{Continuous injection and escape}

In the next step we discuss three more cases where we assume continuous injection and 
escape of the particles. The escape in our calculations may be explained in 
two different ways. First, this term may describe a real escape of the particles into 
the region of a source where the magnetic field strength is significantly smaller 
and therefore the efficiency of the particle emission is also significantly less. 
The escape may also approximate adiabatic losses in the particle energy. 
The description of the adiabatic cooling inside an expanding source is slightly 
different from the description of the escape. However, the precise description of 
the SSC emission of an expanding source is beyond the scope of this paper.

\begin{figure}[!t]
\resizebox{\hsize}{!}{\includegraphics{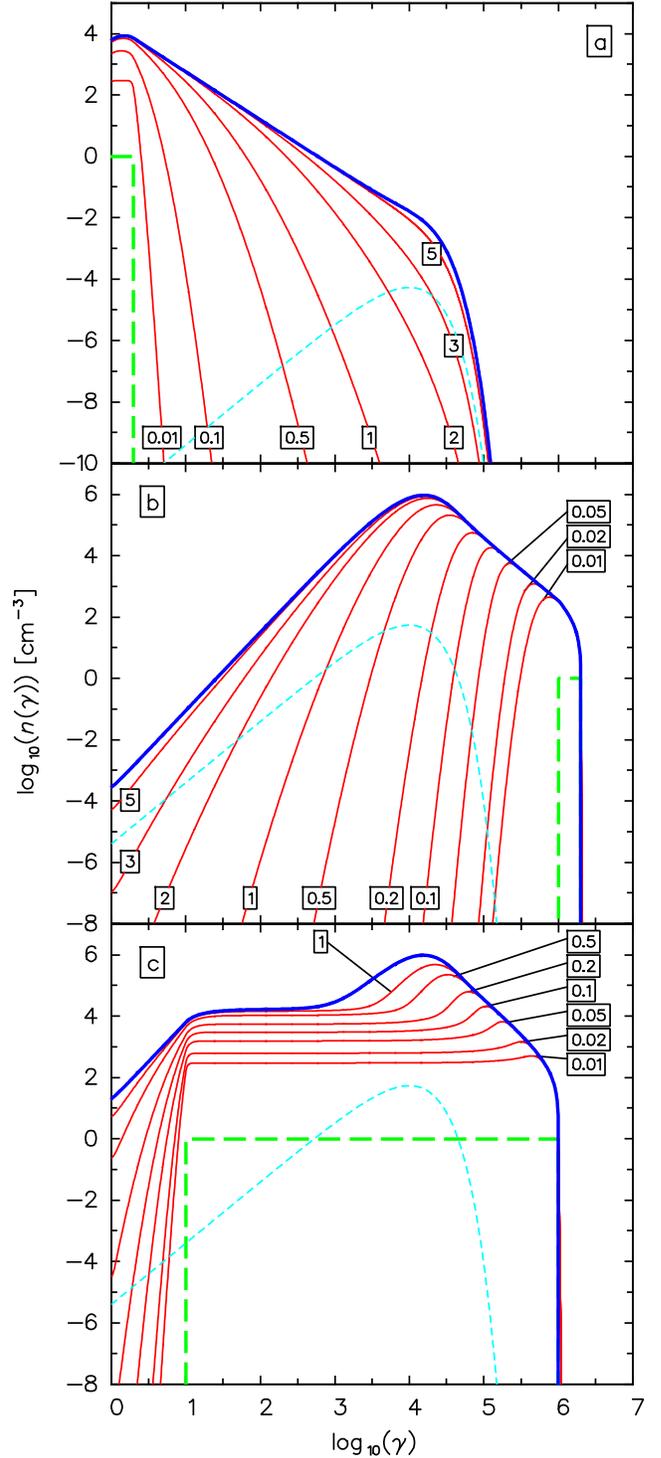}}
\caption{An example of the evolution of the electron spectrum in the case of continuous
         injection and escape of the particles. The bold dashed line shows the 
         initial spectrum and the profile of the injection. We use the same 
         parameters for the initial spectrum and the injection as in the simulations 
         presented in Fig.~\ref{fig_max}. For a comparison, the stationary Maxwellian 
         distributions obtained in the previous tests are presented by a thin 
         dashed line.
        }
\label{fig_pow}        
\end{figure}

In three more tests presented in this subsection we assume that the profiles of 
the continuous injection are equivalent to the initial distributions used in the 
previous tests:

\begin{itemize}
\item[$\bullet$]
     {
      The injection profile in our fourth test is identical to the initial
      distribution in the first test (Fig.~\ref{fig_pow}a). The particles are 
      continuously injected at the lower energy ($1 \le \gamma \le 2$) and 
      systematically accelerated up to the equilibrium energy. The escape of the 
      particles is described by the characteristic escaping time $t_{\rm esc} = 
      t_{\rm acc}$. The competition between the acceleration and the escape in the
      case of continuous injection produces a power law distribution that extends 
      from the maximal energy of the injected particles up to $\gamma_{\rm e}$. 
      Below the maximal injected energy the spectrum is almost flat. Above the 
      equilibrium energy the diffusion process produces an exponential cut--off. 
      This makes this distribution different from the spectrum that could be 
      produced in this case by the deterministic acceleration, where a pile-up
      or a roll-over is produced before the equilibrium energy, depending on whether 
      the particle slope is softer or harder than 2. Moreover, the slope of 
      the spectrum produced by the stochastic acceleration
      \begin{equation}
       n \simeq 1 + \frac{t_{\rm acc}}{2t_{\rm esc}}
       \label{equ_ns}
      \end{equation} 
      differs from the slope that could be obtained from the deterministic acceleration
      \begin{equation}
      n = 1 + \frac{t_{\rm acc}} {t_{\rm esc}}
      \label{equ_nd}
      \end{equation} (e.g. Kirk et al. \cite{Kirk98}). 
      Note that basic shock--acceleration 
      models postulate that $t_{\rm acc} \simeq t_{\rm esc}$, or in more
      sophisticated cases, postulate that $t_{\rm acc}$ and $t_{\rm esc}$ have the same 
      energy dependence. Therefore the index of the particle energy spectrum predicted 
      by such models should be $n \simeq 2$. In our acceleration scenario the value of 
      the characteristic acceleration time is a free parameter, and the resulting 
      index depends on the initial assumption. Moreover, $t_{\rm acc}$ and $t_{\rm esc}$ 
      may have a different energy dependence in a more complex scenario. This may 
      give much more complex spectra. However, as long as the energy dependence of
      the characteristic acceleration and escape time is defined by a power--law 
      function, the resulting energy spectrum should also be a power--law function.
     }
\item[$\bullet$]
     {The continuous injection above the equilibrium energy produces a power
      law spectrum up to the equilibrium. The evolution in this energy range is
      dominated by the cooling and escape process and therefore the index of the 
      produced power law is equal to --2. Below the peak energy, the spectrum is 
      systematically changed from an exponential to a power law at the stationary
      state. This change is similar to the evolution presented in our second test. 
      However, the index of the resulting power law
      \begin{equation}
      n \simeq -2-\frac{t_{\rm acc}}{2t_{\rm esc}},
      \end{equation}
      this time is steeper than the $\gamma^2$ slope necessary to obtain 
      the Maxwellian spectrum in the second test.
     }
\item[$\bullet$]
     {In the last test the particles are injected below and above the equilibrium
      energy (Fig.~\ref{fig_pow}c). The evolution below the minimal injected energy
      is analogous to the evolution in the previous test below $\gamma_{\rm e}$.
      Above the minimal energy, the process is dominated  by the injection and by the
      escape that leads to a flat stationary spectrum. Above the equilibrium energy 
      the evolution is almost the same as in the previous test.
     }
\end{itemize}

The simulations presented in this section show that the injection and escape
may dominate the evolution of the spectrum and may lead to completely different
stationary states. Note that in some of our simulations we assume for the sake 
of simplicity a very low energy of the initial or injected particles 
($1 \le \gamma \le 2$).
However, the turbulent acceleration of such low energetic leptons may be 
difficult in an electron--proton plasma where thermal protons very 
efficiently damps resonant Alfv\`en waves with the frequency equal to their
gyrofrequency. Also a shock wave may have difficulty with the acceleration
of such low energetic leptons. Only the particles with a Larmor radius larger
than the thickness of the shock are able to `feel' the shock discontinuity.
Therefore, some pre--acceleration of the particles is required in order to
let the turbulent or shock acceleration operate efficiently. Such 
pre--acceleration in our test could be simply taken into account by assuming
higher 
energy for the initial or injected particles (e.g. $10 \le \gamma \le 20$). 
Such a relatively small energy increase has a negligible impact on
the evolution of the particle spectrum at very high energies ($\gamma \simeq 
10^6$). Therefore, for the sake of simplicity we neglect the problem of the 
pre--acceleration in the tests, where we assume a relatively low energy for
the initial or injected particles.

\section{Evolution of the SSC emission}

The evolution of the synchrotron emission depends directly on the evolution of
the electron spectrum that also controls the inverse--Compton scattering process.
We analyse here the SSC emission generated by a homogeneous, spherical source
filled by electrons and a uniform magnetic field. The particles are accelerated
uniformly throughout the emitting region. Out of the 6 evolutionary scenarios described
in the previous section we, only investigate here two possible evolutions where 
the initial energy of the particles is relatively low ($1 \le \gamma \le 2$).
Using this simple model we try to explain the high--energy activity of Mrk~501.

\subsection{No injection and escape}
\label{sec_app_noinjesc}

In the first test we assume no injection and escape of the particles from
the acceleration region. 
This simulation is very similar to the first evolutionary 
scenario discussed in Sect. \ref{sec_elecspec}. 

\begin{figure}[!p]
\resizebox{\hsize}{!}{\includegraphics{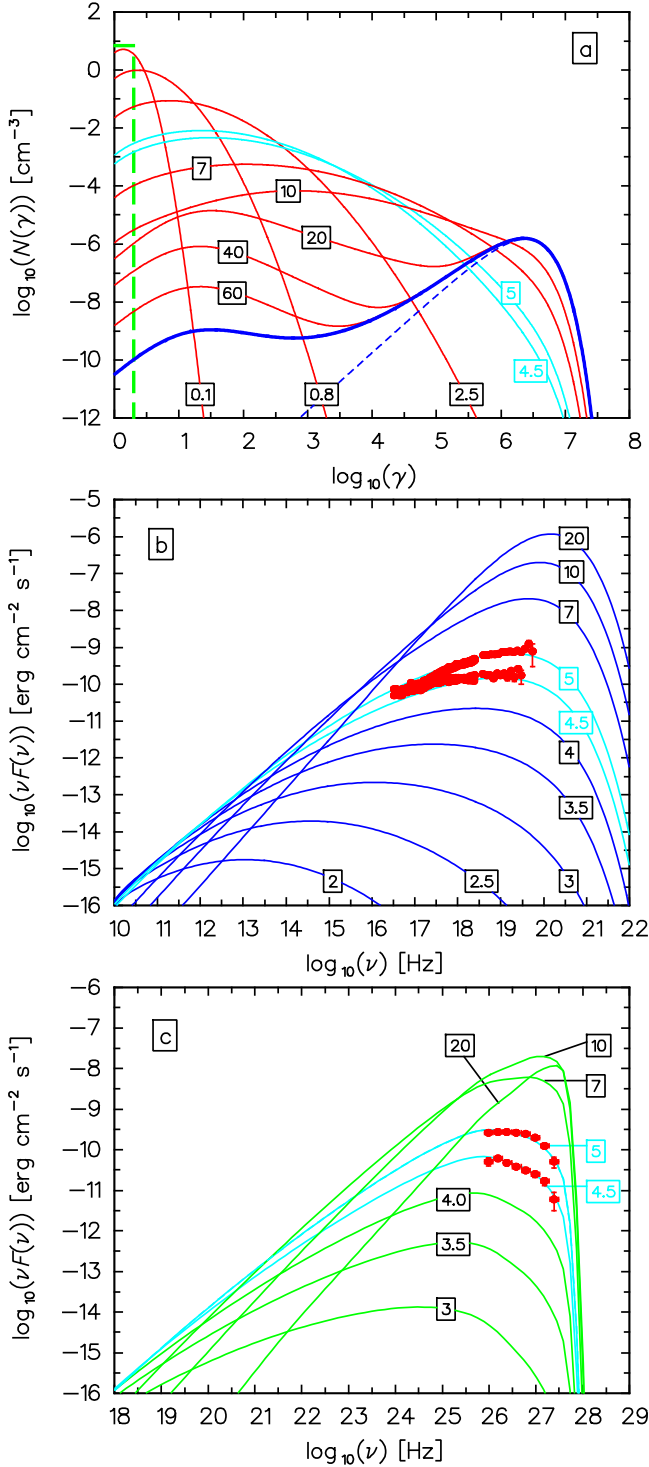}}
\caption{The upper panel (a) shows the evolution of the electron spectrum inside
         a homogeneous, spherical source in the case of no injection and escape
         of the particles. The remaining panels show the synchrotron (b) and the 
         inverse Compton (c) radiation of the source. In this particular test we 
         applied our model to the high energy emission of Mrk~501 observed almost
         simultaneously by the $Beppo$SAX (panel-b, Pian et al. \cite{Pian98}) and CAT 
         (panel-c, Djannati--Atai et al. \cite{Djannati99}) experiments in April 
         1997. The lower level of the emission (7$^{\rm th}$ of April) is 
         reproduced well by the source emission at the evolution time equal to 4.5 
         $t_{\rm acc}$ while the highest level (16$^{\rm th}$ of April) 
         appears to be reproduced well by the emission at the time equal to 
         5 $t_{\rm acc}$.
        }
\label{fig_sscmax}        
\end{figure}

The main difference between
the present simulation and the previous one appears in the particle cooling 
conditions. In the previous test we assumed, for the sake of simplicity, a constant 
cooling ($C = C_0 \gamma^2$), which can be interpreted as the synchrotron emission 
in a constant magnetic field ($U_B \gg U_{\rm rad}$). Here, since we apply our
results directly to the emission of the TeV blazar Mrk~501, we cannot neglect
the radiative cooling due to the IC scattering. Moreover, we have to take
the decrease of the scattering efficiency in the Klein--Nishina
regime into account. This effect is approximated in our calculations by a 
constraint on the radiation--field energy density used for the computation of 
the cooling
\begin{equation}
U_{\rm rad}(\gamma, t) \simeq \frac{4 \pi}{c} 
                       \int_{\nu_{\rm min}}^{\nu_{\rm x}(\gamma)}
                       I_{\rm syn} (\nu, t) d \nu,
\end{equation}
where $I_{\rm syn}$ is the intensity of the synchrotron emission,
$\nu_{\rm x} (\gamma) = \min[\nu_{\rm max}, 3 m_e c^2 / (4 h \gamma)]$,
and $\nu_{\rm min}$, $\nu_{\rm max}$ are the minimal and maximal
frequencies of the synchrotron emission, respectively.  According to
this formula for the very high energetic particles ($\gamma \gtrsim
10^6$), we have $U_{\rm rad} \lesssim U_{\rm B}$ in this particular test.  
On the other hand, the radiation field energy density is much larger than 
$U_{\rm B}$ for the low energy particles ($\gamma \lesssim 10^2$).  
This, as pointed out by Dermer \& Atoyan \cite{Dermer2002}, provides a
quite complex form of the cooling term, which introduces a significant
difference in the evolution of the electron spectrum (see also
Moderski et al. \cite{Moderski05}). This difference is very visible if 
we compare the stationary state in the previous test (Fig.~\ref{fig_max}a) 
and the stationary spectrum in the present simulation
(Fig.~\ref{fig_sscmax}a).  Previously we obtained a pure,
Maxwellian distribution whereas now we have two bumps in the spectrum.
The high--energy bump, that dominates the spectrum appears at the
equilibrium energy related mostly to the acceleration process and the
synchrotron cooling.  The bump that appears at the lower energies is
controlled by the acceleration and the IC cooling that is very strong
only for the low energy particles.  However, as long as the amplitude
of the low energy bump is smaller than the amplitude of the high
energy bump, the existence of the low energy bump is not visible in
the synchrotron spectrum (Fig.~\ref{fig_sscmax}b).  The simple formula
($\alpha = (n-1)/2$) that describes the relation between the electron
spectral index ($n$) and the index of the synchrotron emission
($\alpha$) becomes invalid when $n < 1/3$. In such a case the
spectral index of the power law part of the synchrotron emission
becomes constant and $\alpha=-1/3$.  This value is equivalent to
the spectral index of the synchrotron emissivity of a monoenergetic
population of the electrons, since in this case the electrons
belonging to the high energy bump dominate the total synchrotron
emissivity.

The evolution of the synchrotron radiation presented in Fig.~\ref{fig_sscmax}b 
shows that the $\nu F(\nu)$ peak of the emission is systematically increasing
the amplitude and moving towards higher frequencies. However, the movement 
towards higher frequencies slows down significantly when the maximal particle
energy becomes comparable with the equilibrium energy. On the other hand, the
amplitude increases even if the peak frequency appears almost constant.
The spectral index below the peak frequency changes simultaneously with
the increase in the peak amplitude. However, as discussed above, the
limiting value for the index is $\alpha= -1/3$, and this slope remains
constant even if the peak amplitude is still increasing.

The changes in the IC emission, presented in Fig.~\ref{fig_sscmax}c, are
very similar to the evolution of the synchrotron radiation. However, the
evolution of the peak position appears not to be as strong as in the case of
the synchrotron emission. This is related mostly to the fact that the 
IC scattering is limited by the Klein--Nishina effect. 
Moreover, the very 
high energy gamma rays ($E \gtrsim 1 {\rm TeV}$) are significantly absorbed
due to interaction with the intergalactic infrared radiation field. 
Since we want to apply this simulation to the specific case of Mrk~501,
the spectra shown take into account the absorption suffered by TeV
photons interacting with the cosmic infrared radiation background,
and therefore these cases assume a distance between us and the source
that is equal to the distance of Mrk~501.
The absorption produces a cut--off in the spectra above the frequency $\sim 
10^{28}$~[Hz] in this particular case. To the calculate the absorption,
we use the absorption coefficient derived by Kneiske et al. (\cite{Kneiske02}, 
\cite{Kneiske04}). The spectral index of the IC emission below the peak 
is also limited by the value $\alpha = - 1/3$.

As mentioned above, in order to check whether the discussed scenario 
is able to explain the high energy emission of TeV blazars, 
we apply the results of the simulation to the well--known high energy activity of 
Mrk~501 observed by $Beppo$SAX (Pian et al. \cite{Pian98}) and CAT 
(Djannati--Atai et al. \cite{Djannati99}) in April 1997. 
What is important for these data is that the X--ray/TeV 
observations were made almost simultaneously. However, we have to point out 
that there is a delay of a few days between the higher (16$^{\rm th}$ of April) 
and the lower (7$^{\rm th}$ of April) emission levels and these two emission 
levels were observed during different flaring events (Catanese et al. 
\cite{Catanese97}). Moreover, the X--ray
(as well as TeV) spectra were gathered during a few hours of integration.
The observed variability time scales of TeV blazars at X--ray/TeV energies 
are usually very short, from several minutes up to a few hours. 

Our main aim was to show that our proposed model is able to reproduce 
such very short variability time scales with reasonable assumptions about 
the physical parameters. 
We apply our instantaneous spectra for the observations integrated 
over some period of time to roughly constrain the physical parameters. 
The presented data are one of the best quasi--simultaneous 
observational set ever obtained for TeV blazars. 
We used the same evolutionary scenario 
to explain both presented emission levels.
This means that we use exactly the same values for the physical parameters
in both cases. 
As shown in Fig. \ref{fig_sscmax}b/c, our modeling reproduces the observed 
spectra well if we select the radiation generated at the evolution time 
$t_{\rm evo} = 4.5 t_{\rm acc}$ for the lower emission level and $ t_{\rm evo} 
= 5 t_{\rm acc}$ for the higher level. The characteristic acceleration time 
assumed in the calculations is equal to $R/c$, where $R=3.2\times 10^{15}$ [cm] is the 
source radius. We used also the Doppler factor $\delta=21$, an initial particle 
density $K_{\rm ini} = 7$~[cm$^{-3}$], and the constant magnetic field 
intensity $B=0.05$ [G].

\subsection{Continuous injection and escape}

The second test presented in this section assumes continuous injection and the 
escape of the particles. As already shown (Fig. \ref{fig_pow}a), this scenario
leads to a stationary solution that has a power law shape.

\begin{figure}[!t]
\resizebox{\hsize}{!}{\includegraphics{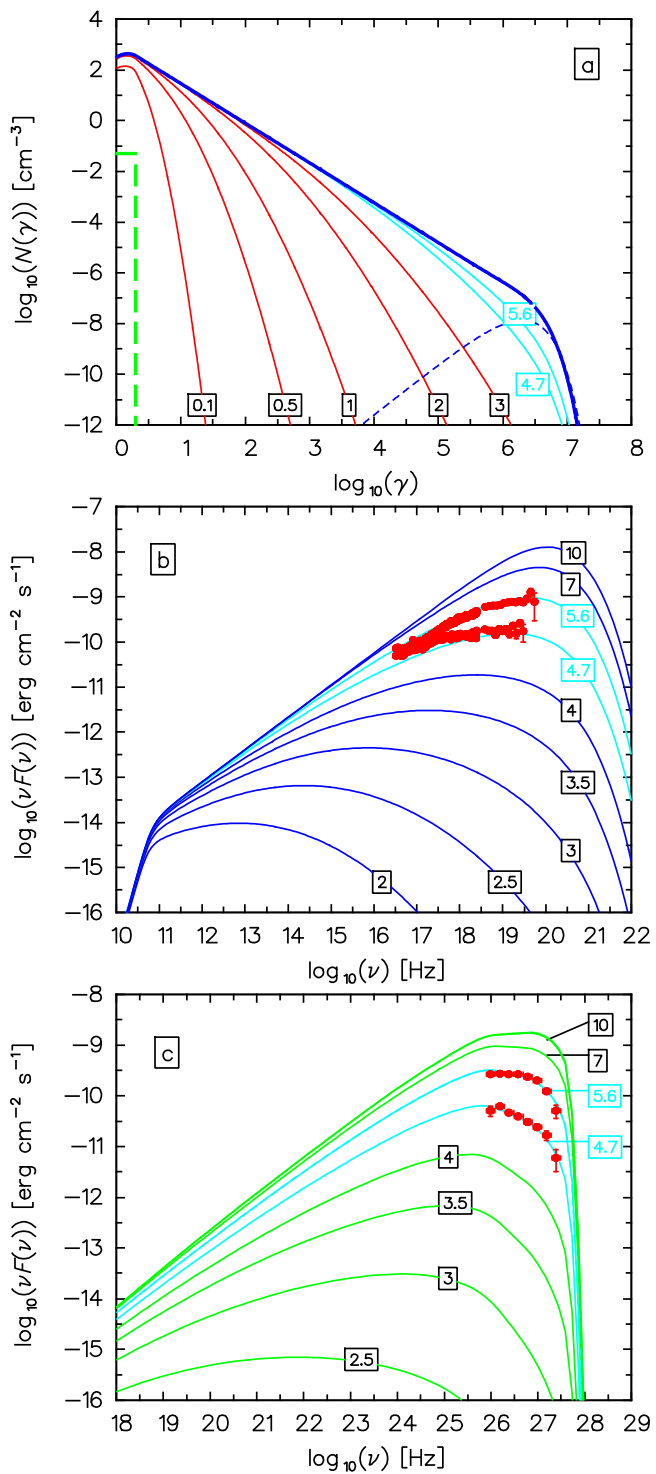}}
\caption{
          The evolution of the electron spectrum (a) and the SSC emission (b, c) in 
          the case of continuous injection and escape of the particles. We apply
          this modeling again for the high energy activity of Mrk~501 observed in
          April 1997 (Pian et al. \cite{Pian98}, Djannati--Atai et al. \cite{Djannati99}).
        }
\label{fig_sscpow}        
\end{figure}

The cooling conditions in the present simulation are similar to the conditions 
described in the previous subsection. However, the complexity of the cooling 
in the Klein--Nishina regime now has no impact on the evolution of the low 
energy part of the spectrum. In the low energy range, the evolution of the 
spectrum is dominated by the acceleration and escape processes.

In the present simulation, the evolution of the peak position is very similar 
to the evolution in the previous test (Fig.~\ref{fig_sscpow}b). The value of the
spectral index below the peak also changes fast with the increase in the
peak amplitude. However, now the limiting spectral index $\alpha \simeq 
0.25$ is related to the assumption $t_{\rm acc} = t_{\rm esc} = R/c$
made in this particular test. The assumption, according to Eq. \ref{equ_ns}, gives 
the spectral index of the energy spectrum $n \simeq 1.5$ and the above--mentioned
index of the synchrotron emission.

If we compare the IC emission in the previous test (Fig.~\ref{fig_sscmax}c) with
the same radiation in the present calculations, we see that the evolution in
both cases is very similar. The scattering in the Klein--Nishina regime 
and the absorption of the TeV emission limit the evolution of the peak.
The difference between the two simulations appears in the slope of the IC 
spectrum below the peak that now is $n \simeq 0.25$. 
The emission level at the end of the simulation (stationary particle
distribution) in the present (as well as in the previous) simulation is 
much higher than the maximal observed level. 
However, the spectra selected 
from the evolution sequence before the stationary state provide a much better 
explanation for the observations than the spectra that correspond to the 
stationary state. 
Moreover, the variability of the TeV sources, where a fast 
increase in the observed flux is followed by an almost equally fast decrease 
(quasi--symmetric flare profile), may indicate that the 
source is almost never able to reach the stationary state. 
Therefore, a significantly higher level of the emission related to the 
stationary particle distribution is rather unlikely. 
Note also that relatively small modification in the high energy ``tail'' 
of the particle distribution produces very strong changes in the 
emission level.

To check the validity of the present simulation we again apply our results 
to the high energy activity of Mrk~501 observed in April 1997.  
We can reproduce the observed spectra well assuming $t_{\rm evo} = 4.7 
t_{\rm acc}$ for the lower emission level and $t_{\rm evo} = 5.6 
t_{\rm acc}$ for the higher level. This may suggest that in 
comparison to the previous test, the system needs twice as much time to
evolve from the lower to the higher emission level. However, the source
radius assumed in the preset test ($R=10^{15}$ cm) is more that three times
smaller than the radius in the previous simulation ($R=3.2 \times 10^{15}$ 
cm). This means that the discussed time was even shorter than in the previous 
case. We use the following values of the other model
parameters: $\delta=33$, $B=0.11$~[G], and the the injected particles 
distribution $Q_{\rm inj}= 5 \times 10^{-2}$  [cm$^{-3}$ s$^{-1}$]. 
The parameters used in this test are quite similar to 
the parameters used in the previous calculations. The specific assumptions about 
some of the parameters and the number of free parameters required by the model 
are discussed in the next section.

\section{Number of free parameters}

In order to explain the SSC emission of a homogeneous spherical source filled
by a uniform magnetic field and monoenergetic population of the electrons, we
have to specify the size of the source ($R$), the magnetic field intensity ($B$), 
the particle density ($K_{\gamma}$), the energy ($\gamma$) of the particles, 
and the Doppler factor ($\delta$) if the source velocity is relativistic. 
This gives only five free parameters. However a monoenergetic population of 
the particles generates a synchrotron spectrum with the index $\alpha = -1/3$ 
and the exponential cut--off that do not provide a good fit to the observed 
spectra of TeV blazars. 

Assuming a power--law distribution of the particle energy, we have to
specify the density ($K_{\rm min}$) of the particles with the minimum energy 
($\gamma_{\rm min} \simeq 1$) instead of $K_{\gamma}$ and the maximal particle 
energy ($\gamma_{\rm max}$) instead of $\gamma$. 
Moreover, we also have to specify the slope ($n$) of the electron distribution. 
This gives one more free parameter in comparison to the first model.
The power--law distribution can explain some of the observations,
but to explain most of the data we have to apply a more complex distribution.

Using a broken power--law energy distribution, we can explain most of the 
observed spectral shapes. 
However, to describe the broken power--law, we have
to specify two more free parameters. This approach requires additional 
information about the break energy ($\gamma_{\rm brk}$) and the slope ($n_2$)
of the second part of the energy spectrum. Therefore the model requires
eight free parameters. It should be mentioned that in some cases when
the second part of the electron spectrum is relatively steep ($n_2 \gg 3$),
the synchrotron and IC peaks are related to the break in the energy 
spectrum. Therefore $\gamma_{\rm max}$ is unimportant, which reduces the number 
of free parameters to seven. 
However, this simplification does not apply
to the high--energy observations of Mrk~501 used in this work. To explain
the data with the broken power law and the simple homogeneous model we
have to specify at least eight free parameters (e.g. Katarzynski et al.
\cite{Katarzynski01})

The model presented in this work uses a more complex spectra than either 
a power or a broken power--law. Our spectra are calculated self--consistently 
according to the assumed physical conditions. Therefore, instead of the 
spectral parameters, we have to describe the processes that control the 
evolution of the spectrum (acceleration, radiative cooling, initial 
spectrum, or continuous injection and possible escape). 
In principle, the model requires
seven free parameters ($\delta$, $R$, $B$, $K_{\rm ini}$ or $Q_{\rm inj}$, 
$t_{\rm acc}$, $t_{\rm esc}$, and $t_{\rm evo}$). However, if we assume 
that the escaping process is negligible, then the number of free parameters
is reduced to six. 
We made such an assumption in the first simulation that
has been applied to the activity of Mrk~501 (Sec. \ref{sec_app_noinjesc}).
Moreover, we assume in this particular test that the acceleration time is
equal to the light crossing time of the source ($R/c$), which reduces the 
free parameters to five. This is a completely arbitrary assumption, but it 
provides reasonable results. To explain the lower of the two observed 
emission levels, the source requires an evolution time equal to 4.5 $R/c$, 
in the first simulation. 
This gives about 6.5 hours in the observer's frame. However,
to get from the lower emission level to the higher level, the source
requires only 0.5 $R/c$, which gives 0.7 h in the observer's frame.
This time scales are in good agreement with the variability time
scales observed in TeV blazars (Maraschi et al. \cite{Maraschi99}, 
Fossati et al. \cite{Fossati00}, Aharonian et al. \cite{Aharonian02}). 

In the second test performed to explain the activity of Mrk~501,
we assumed continuous injection and escape of the particles. The
characteristic escape time was assumed to be equal to the crossing 
time. This assumption describes on average the fastest possible escape
of the particles. Moreover, we assumed that $t_{\rm acc}$ is also equal 
to $R/c$ as in the previous test. 
In fact the acceleration time 
cannot be longer than $R/c$ because the particles could escape from the 
source without a significant energy gain. 
On the other hand, the acceleration time can be shorter than $R/c$, which
could correspond to a more efficient acceleration. 
However, $t_{\rm acc} \ll t_{\rm esc}$ do not provide the required slope 
of the electron spectrum ($n \eqsim 1.5 \dots 2 \to \alpha \eqsim 0.25 
\dots 0.5$). Assuming $t_{\rm acc} = t_{\rm esc} = R/c$ we
reduce the number of free parameters to five ($\delta$, $R$, $B$, 
$Q_{\rm inj}$, $t_{\rm evo}$). 
This gives a significant advantage in 
comparison to a simple homogeneous SSC model with the broken power
law electron spectrum. 
Moreover, our model describes a time--dependent evolution of the source 
emission. This gives the possibility of eliminating one more free 
parameter. Suppose that we have observations of the source in a stationary state. 
Then the evolution time ($t_{\rm evo}$) can be eliminated. 
This reduces the number of free parameters to only four.

The reduction of the number of free parameters with respect to the 
commonly used SSC models may have very important implications. 
Suppose that we can describe the emission with only three free 
parameters. Then the values of these parameters could 
be directly derived from the relationship for the synchrotron peak 
frequency, the level of the synchrotron emission at the peak and the
observed variability time scales. This means that we could predict
the IC emission from the observed X--ray radiation and verify for
example the absorption of the very high energy gamma--rays by the
infrared intergalactic radiation field. 
However, we again have to stress that in constructing our model 
we have made a few completely arbitrary assumptions (e.g. $t_{\rm acc} = R/s$). 
These assumptions provide quite reasonable results but this is not a
proof that our model correctly describes the source evolution. Our 
assumptions require independent observational confirmation. 
Our main aim was only to show how some of the free parameters 
could be eliminated from the model. 
If we neglect the arbitrary assumptions, our time--dependent model 
requires seven independent parameters. This is the number of free 
parameters also required by simple stationary SSC scenario.

\section{Summary}

We have proposed a simple time--dependent model for the SSC emission 
of a homogeneous source, which has direct application for the high energy
emission of TeV blazars. The most important assumption in our modeling
is that the acceleration of the particles has a stochastic nature.
To explain such a case, we use the momentum diffusion equation
that describes the evolution of the particles energy spectrum. For the
sake of simplicity we have not described the possible 
acceleration processes in details but assumed only a constant 
average efficiency of the acceleration that is a free parameter 
in our model.

We first analysed the stationary analytic solution of the diffusion 
equation. We show that for constant cooling conditions and in the 
absence of particle injection or escape, the competition between the
stochastic acceleration and the radiative cooling may lead to a
thermal or quasi thermal distribution of the particles. This result
is in a good agreement with the results of Schlickeiser (\cite{Schlickeiser84}).

In the next step we used numerical solutions of the diffusion equation
to analyse the time--dependent evolution of the electron energy spectrum.
In the case of no injection and escape of the particles, the resulting
stationary spectrum is almost independent of the initial distribution.
Then, assuming a continuous injection and simultaneous escape of the 
particles, we have showed that the stochastic acceleration may lead to 
different energy distributions that depend on the assumptions about the
spectrum of the injected particles. In the most realistic case, when
particles are injected at the lowest possible energies, the stochastic
acceleration in competition with the escaping generates a power--law 
particle distribution.

Finally, we applied our model for the high energy activity of Mrk~501
observed in April 1997.  We have used two opposite scenarios for the
evolution of the electron spectrum.  First we investigated the case
where the escape and injection of the particles is neglected.  In the
second test we instead assumed continuous injection and escape.  In
both cases we have obtained satisfactory fits for the observed
spectra.  We have to point out that our model describes time--dependent
evolution of the emission using the number of free parameters that is
comparable to or even less than the number required by the very simple SSC
scenario.

We point out that a general prediction of this model is that in the
phase of ongoing preequilibrium stochastic acceleration of a population
of low--energy electrons (see Fig. \ref{fig_sscmax}/\ref{fig_sscpow} a,b),
the emissivity level of the peak of the synchrotron emission and the energy 
of the peak are correlated, a behaviour already observed a few times in 
blazars (e.g. Fossati et al. \cite{Fossati00}, Tanihata et al. 
\cite{Tanihata04}). We are going to investigate this problem in detail in 
our further work. We will also investigate some specific relations between 
the X-ray and the TeV light curves that are predicted by the model.

Finally, a straightforward extension of our study is the application to the
case of powerful blazars, in which the dominant role in the cooling of
the high-energy electrons is played by the external radiation field of
the host quasar. In these sources, as opposed to the case of TeV
BL Lacs, the low-energy branch of the IC component is well known from
X-ray observations, providing a supplementary constraint to the model.

\begin{acknowledgements}
We thank the anonymous referee for a number of constructive comments that
improved the paper. We are grateful to E. Pian, and A. Djannati--Atai, for 
the data obtained by the $Beppo$SAX and CAT experiments. We acknowledge 
the EC funding under contract HPRCN-CT-2002-00321 (ENIGMA network).
\end{acknowledgements}

\end{document}